\documentclass[aps,prb,reprint,superscriptaddress,showpacs]{revtex4-1}
\usepackage{color}
\usepackage{graphicx}
\usepackage[american]{babel}
\usepackage[T1]{fontenc}
\usepackage{amsmath}
\usepackage{amssymb}
\usepackage{amstext}
\usepackage{amsthm}
\usepackage{latexsym}
\usepackage{verbatim}
\usepackage{natbib}
\usepackage{array}
\usepackage{color} 

\usepackage{ulem}   

\def\214{Sr$_2$IrO$_4$}
\def\327{Sr$_3$Ir$_2$O$_7$}

\begin{document}

\title {Coherent and incoherent bands in La and Rh doped Sr$_3$Ir$_2$O$_7$}
\vspace{1cm}

\author{V. Brouet}
\author{L. Serrier-Garcia}
\author{A. Louat}
\author{L. Fruchter}
\affiliation {Laboratoire de Physique des Solides, CNRS, Univ. Paris-Sud, Universit\'{e} Paris-Saclay, 91405 Orsay Cedex, France}

\author{F. Bertran}
\author{P. Le F\`evre}
\author{J. Rault}
\affiliation {Synchrotron SOLEIL, L'Orme des Merisiers, Saint-Aubin-BP 48, 91192 Gif sur Yvette, France}

\author{A. Forget}
\author{D. Colson}
\affiliation{Service  de  Physique  de  l'\'{e}tat  Condens\'{e},  DSM/IRAMIS/SPEC, UMR  3680  CNRS, CEA  Saclay  91191  Gif  sur  Yvette  cedex  France}

\begin{abstract}

In \214 and \327, correlations, magnetism and spin-orbit coupling compete on similar energy scales, creating a new context to study metal-insulator transitions (MIT). We use here Angle-Resolved photoemission to investigate the MIT as a function of hole and electron doping in \327, obtained respectively by Ir/Rh and Sr/La substitutions. We show that there is a clear reduction as a function of doping of the gap between a lower and upper band on both sides of the Fermi level, from 0.2eV to 0.05eV. Although these two bands have a counterpart in band structure calculations, they are characterized by a very different degree of coherence. The upper band exhibits clear quasiparticle peaks, while the lower band is very broad and loses weight as a function of doping. Moreover, their ARPES spectral weights obey different periodicities, reinforcing the idea of their different nature. We argue that a very similar situation occurs in \214 and conclude that the physics of the two families is essentially the same. 

\end{abstract}

\date{\today}

\maketitle
\section{Introduction}

Metal-insulator transitions are a central feature of correlated systems \cite{ImadaRMP98,GeorgesRMP96}. They are often driven by doping, which is particularly well adapted to angle resolved photoemission (ARPES) studies, but still difficult to describe theoretically. Should the gap close progressively ? Will all carriers contribute to the emerging Fermi Surface (FS) or just those added by the dopants ? For a correlated system, coherent quasiparticles bands are expected to form within the gap, coexisting with incoherent Hubbard bands on both sides of the gap \cite{GeorgesRMP96}. These bands, as well as possible transfer of spectral weight between them, can be observed by ARPES. Consequently, this technique has been used extensively for the study of cuprates, where a particularly complicated situation takes place, as the FS emerges through a pseudogap phase, where parts of the FS are partially gapped \cite{VishikCondmat18}. Deciding whether this is intrinsic to cuprates or more generic awaits the investigation of more types of Mott insulators. 

\begin{figure}[tbh]
\centering
\includegraphics[width=0.5\textwidth]{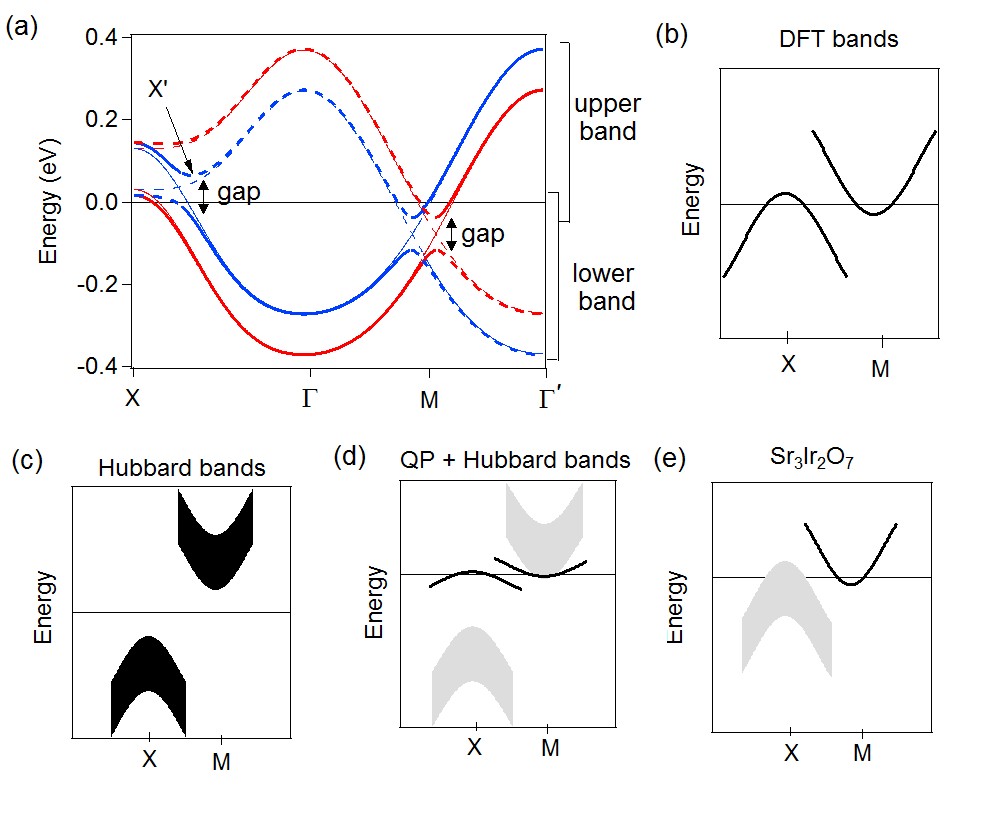}
\caption{(a) Sketch of the band structure expected for \327 in DFT calculations (thick lines, see appendix for full calculation). Two bands cross the Fermi level that we call lower and upper band, forming respectively hole pockets at X and electron pockets at M. They can be viewed as originating from the thin lines with a gap (see arrow) opening at their crossings. The thin lines are similar to the J=1/2 band of \214 (see appendix), doubled because of the bilayer. The colors refer to the two bands of the bilayer and the solid/dotted lines to the direct/folded bands due to the 2 Ir in-plane (see appendix). (b) Simplified version of (a). (c) Same as (a) with a larger gap between lower and upper band. The two bands could be viewed as DFT bands split by a large magnetic gap, as obtained for example in a magnetic LDA+U calculation. They could also be viewed as incoherent Hubbard bands of a DMFT calculation. They are shown here shifted in k to recall the initial semi-metallic situation, but this is not an important point. The larger width of the lines is meant to symbolize the incoherent nature. (d) Scenario expected for electron doping of a Mott insulator : the Hubbard bands do not shift significantly, but transfer weight (their lower weight is symbolized by the grey color) to a QP band (solid lines) forming at the bottom of the upper Hubbard band. This QP band is typically strongly renormalized with respect to DFT bands. (e) What we observe in \327 seems an intermediate situation, where the upper band behaves as a DFT or weakly renormalized QP band and the lower band as a shifted Hubbard band. } 
\label{Fig_Sup_327}
\end{figure}

Layered iridates offer a new test family, which is unusual as correlations take place in presence of strong spin-orbit coupling (SOC). In \214, the SOC is necessary to remove the degeneracy of the ground state and form a narrow half-filled band at the Fermi level, where modest correlations open a Mott gap \cite{BJKimPRL08,MartinsPRL11,AritaPRL12}. Recent theoretical studies discuss whether, upon doping, the transition to the metallic state should occur through a pseudogapped phase \cite{MartinsPRM18,MoutenetPRB18}. \327 shares many properties with \214 \cite{CaoPRB02,DhitalPRB13}, but the role of correlations is not as obvious. Its bilayer structure creates a band gap of about 0.1eV in the doubled J=1/2 bands near the Fermi level, even without correlations \cite{CarterPRB13_TB,OkadaNatMat13}). Fig. \ref{Fig_Sup_327} sketches this gap opening, which results in a semi-metallic case in density functional theory (DFT) band structure calculations (full calculations are shown in appendix). Although correlations and/or magnetism are usually found necessary to enlarge this gap and reach the insulating state \cite{CarterPRB13_TB,OkadaNatMat13}, the two states are adiabatically connected, which is different from \214 \cite{DeLaTorrePRL14}. Indeed, one ARPES study of \327 doped with La concluded that it should be viewed as a doped semiconductor, very different from \214 and with little traces of correlations \cite{DeLaTorrePRL14}. A subsequent ARPES study clarified that there is nevertheless a gap that closes as a function of doping \cite{HeNatMat15} and STM \cite{OkadaNatMat13,HoganPRL15} or optical studies \cite{MoonPRB09,AhnSciRep16} also support a correlated behavior. As it would be quite interesting to have two examples of \lq\lq{}spin-orbit Mott insulators\rq\rq{}, with different strength of correlations and different symmetry of magnetic excitations\cite{KimPRL12_327}, understanding how these two systems should be compared is crucial. 

To address these questions, we present an ARPES study of \327 doped with La or Rh. To our knowledge Rh doping, which substitutes for Ir, was never reported in \327. We show it induces effective hole doping as it does in \214 \cite{ClancyPRB14,QiCaoPRB12,CaoNatCom16,LouatPRB18}. Previous ARPES studies have well documented the emerging FS under high La doping, characterized by small \lq\lq{}lens-like\rq\rq{} electron pockets around the M point \cite{DeLaTorrePRL14,HeSciRep15,HeNatMat15}. We focus on the bands below the Fermi level to determine more completely the evolution of the electronic structure. We confirm the reduction of the gap, not only at the X point of the Brillouin Zone (BZ) as previously reported \cite{HeNatMat15}, but also at the M point. At first sight, this brings the system back to the non-correlated case. Our key finding is however that, beyond the relative position of the bands, their nature is very different, as sketched in Fig. 1e. The band at the Fermi level exhibits well defined quasiparticle peaks, while the lower band has a much larger linewidth and loses weight as a function of doping, as expected for an incoherent band. This points to the presence of correlation effects, which go beyond predictions of simple band calculations. We also uncover different intensity modulation of these bands over neighboring BZ. The coherent band spectral weight follows the periodicity of the true unit cell (containing two Ir \cite{HoganPRB16}), while the incoherent band is mainly sensitive to on-site properties. We conclude by a comparison with \214, where a similar dichotomy between the upper and lower bands can be deciphered.

\section{Experimental details}

Single crystals of (Sr$_{1-x}$La$_x$)$_3$Ir$_2$O$_7$  with x=0, 0.016, 0.06 and Sr$_3$(Ir$_{1-x}$Rh$_x$)$_2$O$_7$ with x= 0.03 were grown as follows. High purity powders of SrCO$_3$ (99.995\%), IrO$_2$ (99\%) (La$_2$O$_3$ (99.999\%) or Rh$_2$O$_3$ (99.9\%)) were dried, weighed, mixed in a glove box under argon with SrCl$_2$ (99.5\%) flux in the ratio 3:2:5. The mixture was loaded into a platinum crucible covered with a platinum tip, reacted in tubular furnace under oxygen flux (except for Rh, synthesized in air) at 1120$^\circ$C for 6 hours and slowly cooled (10$^\circ$C/h) to 600$^\circ$C. Then deionized water was used to dissolve the SrCl$_2$ flux and extract the single crystals. The crystals are platelets with the smallest dimension along [001] direction and 0.3 to 1mm as side. The exact doping was estimated by Energy Dispersion X-ray analysis.  Unit cell dimensions of pure  \327 have been determined by X-ray diffraction data collected at 150 K on a Nonius Kappa-CCD area detector diffractometer using graphite-monochromated Mo K$\alpha$ radiation ($\lambda$ = 0.71073 \AA) : a = b= 3.8951(2) \AA~and c = 20.8941(13) \AA~with  tetragonal space group I4/mmm. No impurity phases were detected for the pure and La cases, but about 5\% pure \214 was detected by SQUID measurements in the Rh case. However, ARPES signature of \214 is easy to distinguish from \327 and none was observed on the cleaved surface for our measurements. Typical resistivity curves are shown in appendix. They are similar to those published in literature for the same La dopings \cite{LiCaoPRB13,HoganPRL15}. For 3\% Rh, the resistivity is in between the two La cases, on the insulating side of the transition. It increases weakly with decreasing temperatures. 

ARPES experiments were carried out at the CASSIOPEE beamline of SOLEIL synchrotron, with a SCIENTA R-4000 analyser and an overall resolution better than 15~meV. All data shown here were acquired at a photon energy of 100~eV, with linear polarization along $\Gamma$M. The temperature is 50~K, which is in the magnetic phase for all compounds, except for 6\% La doping, where no magnetic transition is detected by SQUID measurements.
\begin{figure}[tb]
\centering
\includegraphics[width=0.5\textwidth]{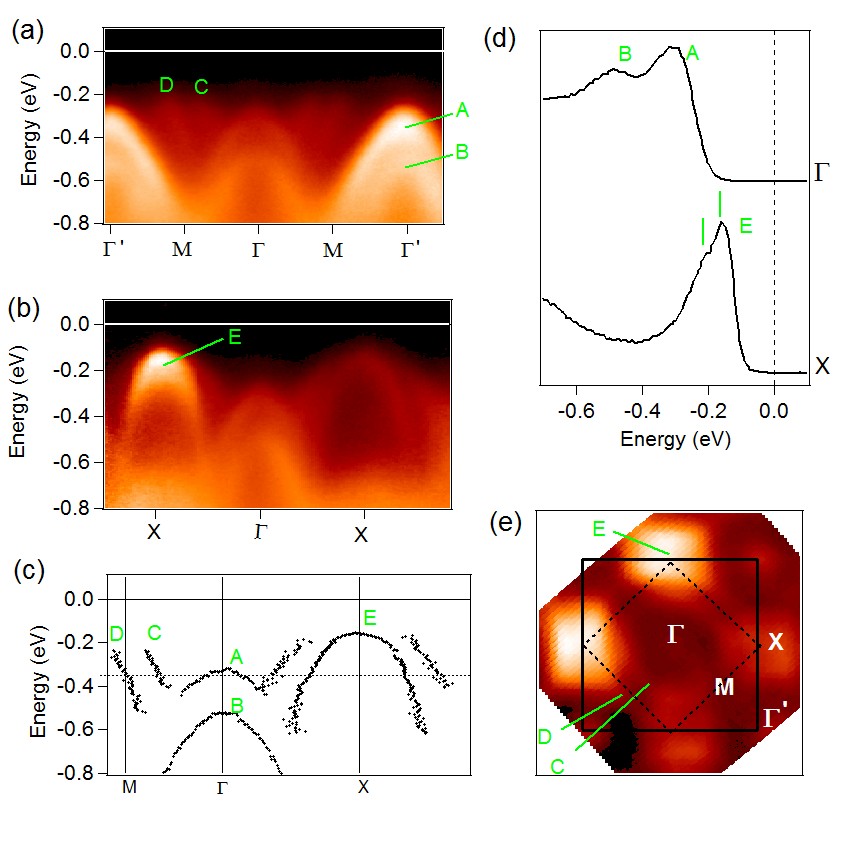}
\caption{Energy-momentum plots of the dispersion along $\Gamma$M (a) and $\Gamma$X (b) in \327. (c) Dispersions extracted from (a) and (b). (d) EDC peaks at $\Gamma$ and X. (e) Map of the spectral weight integrated at -200meV in a 20meV window. The thick black square delimits the 1Ir BZ and the dotted thiner black square the 2 Ir BZ. }
\label{Fig_DatasPure}
\end{figure}


\section{ARPES in S\lowercase{r}$_3$I\lowercase{r}$_2$O$_7$}

\begin{figure*}[tbh]
\centering
\includegraphics[width=0.95\textwidth]{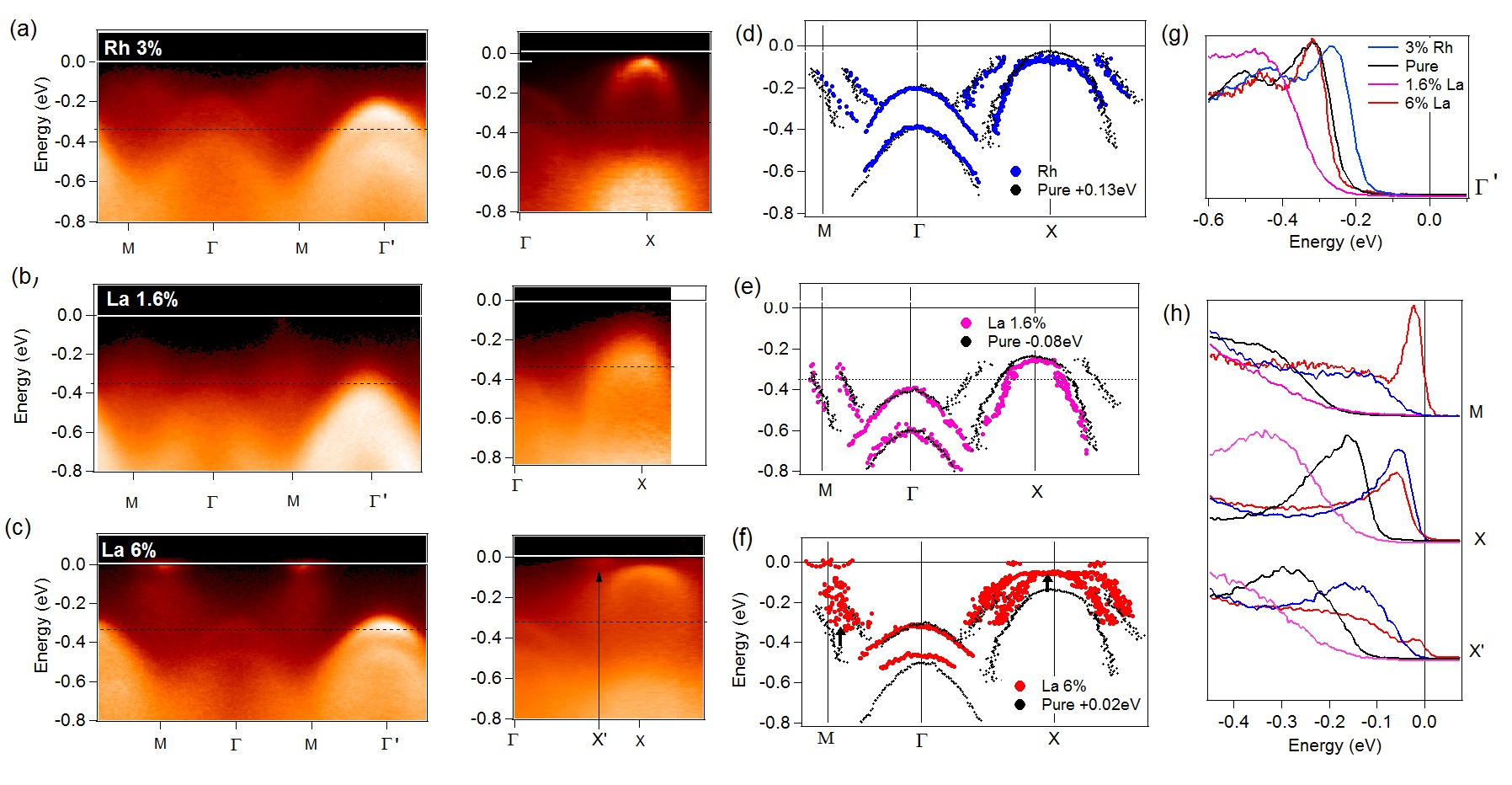}
\caption{ (a-c) Energy-momentum plots of the dispersion along $\Gamma$M and around X in \327 doped with 3\% Rh (a), 1.6\% La (b) and 6\% La (c). The black dotted line indicates the position of the J=3/2 band in \327. (d-f) Extracted dispersion, compared with that of the pure (Fig. \ref{Fig_DatasPure}), shifted as indicated. (g) EDC at $\Gamma$ compared between selected samples. (h) EDC at M, X and X\rq{} (X\rq{} is defined in panel c).}
\label{Fig_Doped}
\end{figure*}
We start with a discussion of the band structure of \327 to set clear the context of our measurements. In Fig.\ref{Fig_DatasPure}, we show the energy-momentum plots of ARPES intensity obtained in \327 along $\Gamma$M (a) and $\Gamma$X (b). Similar results were published previously \cite{WangDessauPRB13,KingPRB13,MoreschiniPRB14}. At $\Gamma$, the characteristic double peak structure of the J=3/2 band can be recognized. The two J=3/2 doublets are completely filled with 4 electrons. The two peaks at -0.35 and -0.5eV are noted A and B and are also shown in the EDC at $\Gamma$ in Fig.\ref{Fig_DatasPure}d. There is a clear alternation of weak bands at $\Gamma$ and strong bands at $\Gamma$\rq{} for this double peak. These two points are equivalent in a 2 Ir BZ (dotted square in Fig. \ref{Fig_DatasPure}e) but inequivalent in a 1Ir BZ (solid square). Hence, the relative intensities are modulated with a strength proportional to that of the symmetry breaking \cite{VoitScience00}. 

In Fig.\ref{Fig_DatasPure}a, we also see quite clearly bands going up from $\Gamma$ to M with a maximum around -0.2eV at C and D. Their dispersion is reported in Fig. \ref{Fig_DatasPure}c. They have not been discussed in previous ARPES papers, but they form two well defined circles around $\Gamma$, as can be observed in Fig. \ref{Fig_DatasPure}e with the map of the ARPES spectral weight integrated at -0.2eV. They evidently correspond to the J=1/2 band gapped by $\sim$0.2eV. Their map reminds the circular FS expected for the J=1/2 band in \214 (with an additional splitting) and their dispersion corresponds well to the one of the J=1/2 band (a basic band structure of \214 is recalled in appendix). Only the part from $\Gamma$ to M is clearly visible in ARPES, because the other side is folded in the 2Ir BZ and is then very weak, as for the J=3/2 bands at $\Gamma$. This explains the anomalous \lq\lq{}stopping\rq\rq{} at C and D, also observed in \214 (see appendix). 

The J=1/2 band is also clear at X, with a maximum located at E, near -0.2eV (Fig. \ref{Fig_DatasPure}b and \ref{Fig_DatasPure}d for the corresponding EDC). We note that there seems to be a small splitting of the band at X, of about 0.1eV (see marks on the EDC), which may either be an intrinsic structure of the band due to correlations or may reveal a further lowering of symmetry. 

\section{ARPES in doped S\lowercase{r}$_3$I\lowercase{r}$_2$O$_7$}

Fig. \ref{Fig_Doped} displays the evolution of the electronic structure for Rh (a) and La (b-c) dopings. At first sight, they look very similar to the pure case, except for 6\% La, where small pockets appear at the Fermi level. There are nevertheless significant shifts of the structures for the low dopings. As a reference, we show by black dotted line the position of the first J=3/2 band in \327 (-0.35eV). For 3\% Rh doping (Fig.\ref{Fig_Doped}a), all bands move up by 0.13eV, which is confirmed by the nearly perfect overlap of the dispersions when this shift is applied, as done in Fig.\ref{Fig_Doped}d. Although the bands at X are very close to E$_F$, the leading edge is still at -20meV (see Fig. \ref{Fig_Doped}h), in agreement with the non-metallic character. Also, the double peak structure that existed in the pure case at X has weakened or disappeared. This rigid shift up towards the Fermi level is analogous to the one observed in \214 and can be attributed to effective hole doping \cite{CaoNatCom16,LouatPRB18}. 

For 1.6\% La (Fig. \ref{Fig_Doped}b), there is also a rigid shift, except it is of -0.08eV, to higher binding energies, consistent with electron doping. The bands also tend to broaden, which may be due to inhomogeneities introduced by doping. As can be seen on the EDC spectra in Fig. \ref{Fig_Doped}h, there are no in-gapped states around -0.1 eV, neither at M nor X, contrary to what was reported in \cite{AffeldtPRB18}. Such states then probably comes from some type of disorder and/or inhomogeneities. We also do not observe any significant change of the effective mass for the main bands, as can be seen from the good overlap of all dispersions, contrary to what was reported based on these in-gapped states \cite{AffeldtPRB18}. 

It is likely that the Fermi level is very near the top of the lower band for 3\% Rh and the bottom of the upper band for 1.6\% La (see Fig. 1 for a definition of what we call lower and upper bands). This gives an estimate of the gap of about 0.2eV, in good agreement with other measurements, like STM \cite{HoganPRL15}. We have found a very similar behavior in \214 \cite{BrouetPRB15}, so that this shift of Fermi level characterizes the low doping behaviors in these iridates. We note that for the pure compound, E$_F$ is closer to the electron doped side. This may indicate that our sample is slightly electron doped. Indeed, bands around -0.25 and -0.45eV were reported in literature of \327 \cite{MoreschiniPRB14,KingPRB13}, 50meV above ours, and shifting to lower values with La doping.

\vspace{0.25cm}

On the contrary, in 6\% La [Fig. \ref{Fig_Doped}(c) and (f)], the shift is not rigid anymore. The J=3/2 band is near the position of the pure case, but, remarkably, the J=1/2 bands at X have moved up by 0.15eV (see arrows). This evidences a true reduction of the gap within the lower and upper J=1/2 bands. As we estimated the gap in the pure case to 0.2eV, this reduction is very sizable. The remaining gap of about 50meV is even smaller than the \lq\lq{}structural gap\rq\rq{} obtained in band structure calculations (about 100meV, see appendix and ref. \cite{CarterPRB13_TB,OkadaNatMat13}), so that it seems the magnetic/correlated part of the gap has totally collapsed. We do not know if this smaller gap should be assigned to renormalization effects or shortcomings of the band structure calculations. An additional confirmation of this gap reduction is the appearance of the upper band at the Fermi level, forming small pockets at the position X\rq{} (arrow in Fig. \ref{Fig_Doped}c, see also Fig. 1a). In fact, the position of the lower band at X is almost the same as for Rh, but this pocket is not present in the Rh case, in good agreement with the idea that there is still a larger gap for Rh. A small QP peak appears at X', which can be better seen on the spectra in Fig. \ref{Fig_Doped}h. Importantly, detecting a QP there, even if it is small, means that there is no \lq\lq{}pseudogap\rq\rq{} in this region, which corresponds to the \lq\lq{}antinodal region\rq\rq{} in cuprates and is the one where a pseudogap is discussed in \214 for similar La dopings \cite{DeLaTorrePRL15} or surface doping \cite{KimScience14}. 

The same situation is observed at M. The upper band forms small pockets near E$_F$, as reported before \cite{DeLaTorrePRL14,HeNatMat15}, but we also see trace of the lower band shifted up by 0.15eV (see Fig. \ref{Fig_Doped}c). Its dispersion becomes however quite diffuse and difficult to track, as evidenced by the larger scattering in experimental points. This loss of coherence is anomalous in a band picture. Indeed, this hump was previously identified as the incoherent part of the QP in ref. \cite{DeLaTorrePRL14}. The connection with the lower J=1/2 band questions its nature (this connection was noted in ref. \cite{DeLaTorrePhD}). An EDC spectrum at M in Fig. \ref{Fig_Doped}(h) shows that these lower bands form a hump of full width $\sim$160meV, below the QP peak of full width $\sim$40meV. Similarly, the peak due to the lower J=1/2 at X\rq{} has evolved into a hump in the 6\% La case. It is much broader and has smaller weight than in the pure and Rh-doped cases. The much larger width is not simply due to the higher binding energy, as the peak in J=3/2 remains relatively narrow despite its higher binding energy (full width $\sim$80meV, see Fig. \ref{Fig_Doped}g). The difference of behavior in width and weight between the two bands is surprising because they have similar orbital character. In a band view, there should be no difference between the upper and lower band of J=1/2, contrary to what we see in Fig. \ref{Fig_Doped}(h). The change in spectral weight rather reminds a correlated case. Evidently these two bands have a different \lq\lq{}status\rq\rq{}, which will be discussed further in conclusion. 

\section{Intensity modulations}
  
\begin{figure*}[tbh]
\centering
\includegraphics[width=1\textwidth]{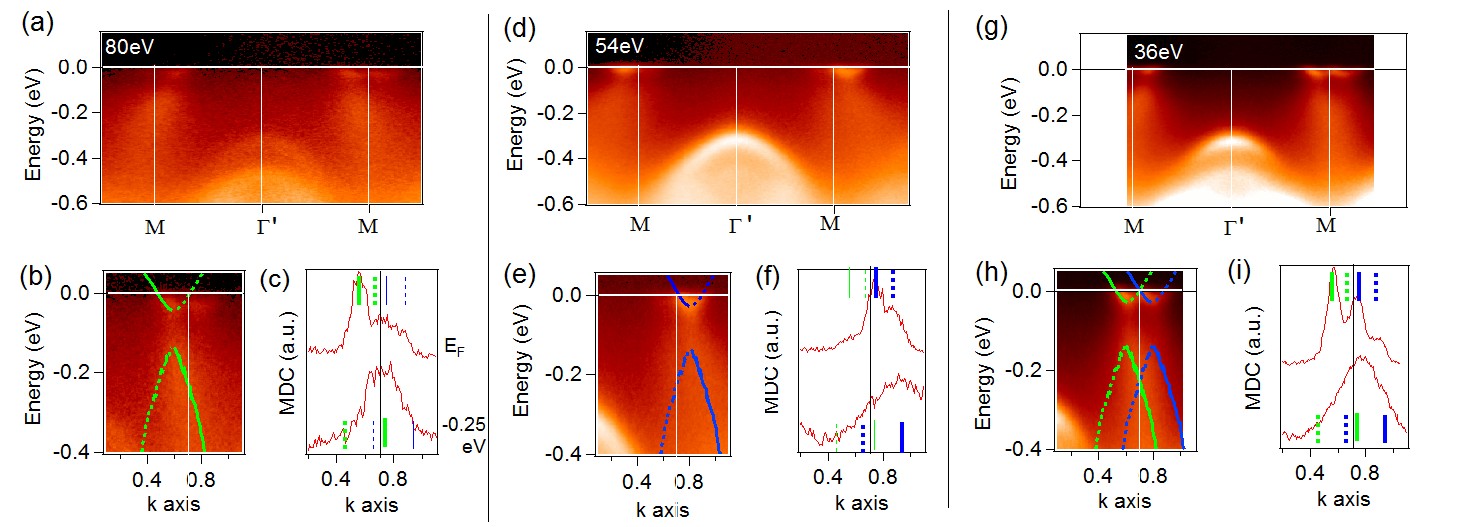}
\caption{(a,d,g) Energy-momentum plots of the dispersion along $\Gamma$\rq{}M in sample with 6\% La, at photon energies of (a) 80, (d) 54 and (g) 36eV, corresponding respectively to k$_z$=1, 0, 0.5. (b,e,h) Zoom on the dispersion around M (k=0.707) of the image above. Thick lines : Sketch of the J=1/2 bands expected around M. The different colors correspond to bilayer splitting and the solid (dotted) line to the direct (folded) character. (c,f,i) MDC spectra at E$_F$ (top) and -250meV (bottom), showing the oscillation of weight between the different bands. The vertical markers indicate the positions expected for the four different bands.}
\label{Fig_VsE}
\end{figure*}
Another difference between the lower and upper bands is detailed in Fig. \ref{Fig_VsE}. While the pockets at E$_F$ are rather symmetric, it is never the case for the lower tail. Only the part of the dispersion going up from $\Gamma$ to M is clear, its "folded" side (dotted line in Fig. \ref{Fig_VsE}) is simply missing.

The green and blue bands on this figure correspond to the bonding and antibonding bands formed by the two layers. The change of phase within the bilayer gives rise to an oscillation of intensity between green and blue bands as a function of k$_z$. This can be probed by changing the photon energy and was observed in Bi cuprates \cite{KordyukPRL02} or \327 \cite{MoreschiniPRB14}. This can be conveniently observed at the $\Gamma$ point between the two J=3/2 bands. At 80eV (Fig. \ref{Fig_VsE}a), the weight is almost entirely in the band at -0.5eV; at 54eV, it is in the one at -0.3eV and at 36eV, it is similar in both. The J=1/2 bands near M follow exactly the same trend, both for the small pocket at E$_F$ and the tail. The left one is strong for 80eV, the right one for 54eV and they have similar weight at 36eV. This can also be seen by the momentum distribution curves of Fig. \ref{Fig_VsE}. The vertical lines indicate where the peaks are expected for the 4 different bands and it is easy to check that it is almost entirely on the green ones at 80 eV, the blue ones at 54 eV and is more evenly distributed for 36eV, both for the upper and lower bands. 

The dephasing between the 2 Ir in the plane gives rise to a very similar modulation of intensity, but this time it is in-plane, between the direct and folded bands (solid and dotted lines). This is very obvious between $\Gamma$ and $\Gamma$\rq{} for J=3/2, as we discussed before with Fig. \ref{Fig_DatasPure}a. As for the J=1/2 band, the same modulation is observed for the lower band (the part from $\Gamma$ to M is strong, but the part from M to $\Gamma$\rq{} is not visible), but not for the upper band forming quite a symmetric pocket. This can be appreciated more quantitatively on the MDC spectra of Fig. \ref{Fig_VsE}. Two peaks can be distinguished on solid and dotted lines at $E_F$, while the spectra are mostly peaked on the solid lines at 250meV. Although it was shown \cite{HeSciRep15} that, in some conditions, one can also obtain an asymmetric pocket (which changes the appearance of the FS from a pocket to a Fermi arc), it remains that the modulation of intensity is much more ubiquitous for the lower band. 

As the intensity of the folded bands is related to the strength of the potential at the origin of the symmetry breaking, this suggests that the upper band probes a larger difference between the two Ir than the lower one. Following our previous observations, we could suggest that the lower/incoherent band is essentially sensitive to on-site properties (hence obeying a 1Ir BZ periodicity), while the upper/coherent band delocalizes over many sites and therefore respects the true 2Ir BZ periodicity of the unit cell.
 
Interestingly, a very similar behavior was observed for \214 at high La doping \cite{DeLaTorrePRL15,GretarssonPRL16}. A clear and symmetric pocket is observed at M near E$_F$, followed by a tail, which rapidly loses weight and is not the symmetric part of the Dirac cone expected in the calculation. On the other hand, when the lower J=1/2 band is clear, for smaller dopings, only the direct part is observed, as shown in appendix.

\section{Conclusion}

Our study points to a large role of correlations in \327. Even though it has a structure that already breaks the J=1/2 doublet into lower and upper bands, the size of the gap between them changes as a function of doping, from 200meV to 50meV, demonstrating the impact of correlations and/or magnetism. Furthermore, \327 presents a peculiar type of insulator to metal transition, where the bands have intermediate character between \lq\lq{}Hubbard\rq\rq{} and \lq\lq{}coherent\rq\rq{} bands. On the one hand, we observe for the lower band spectral weight transfer and large linewidth that are typical of Hubbard bands. On the other hand, this band exists in a DFT calculation, suggesting a coherent nature. This situation may be closer to that of an antiferromagnetic insulator, where DMFT \cite{CamjayiPRB06} predicts \lq\lq{}Slater\rq\rq{} bands to appear within the Mott gap, which may have such an intermediate status. 

The bands we observe in \327 are not renormalized compared with DFT calculations, which is usually taken as a sign of small correlations. However, correlations are patent in the width and loss of weight behaviors of the lower band. The absence of renormalization should rather be viewed here as a sign of the incoherent nature of the bands away from $E_F$. The coherent region in the vicinity of the Fermi level is very narrow, of the order of 50 meV. If this is the energy scale for coherence in this system, it is indeed a rather strongly correlated metal. 

Our study of \327 also gives a fresh view on the situation in \214. As a gap is always present in \327, it is possible to study independently the distance between the lower and upper bands and their respective widths. This makes the separation of coherent and incoherent bands on the two sides of the gap easier. In \214, a clear and symmetric pocket can be observed at the Fermi level\cite{DeLaTorrePRL15,GretarssonPRL16} followed by an incoherent tail. It is difficult to determine whether the gap is closed or not, as the incoherent tail of the lower band extends to the upper band, yielding a very asymmetric shape. This asymmetry is precisely the one we have described between the lower and upper bands of \327, suggesting a unified picture. This has important consequences to discuss the presence/absence of a pseudogap. In \327, the FS should always consist of pockets around M and X and we have shown that when a peak appears at X, it is not gapped. In \214, a simple circular FS would be expected if the gap is closed, at variance with the pocket observed at intermediate dopings, both for La doping \cite{DeLaTorrePRL14} and surface doping \cite{KimScience14}. If the gap is really completely closed, observing a pocket means a pseudogap is present, but if it is only partially closed, it is natural to observe only a pocket around M. A progressive closing of the gap is expected from the behavior of \327 \cite{HeNatMat15}, so that the meaning of these pockets is not straightforward. On the other hand, the different width of the lower and upper bands emerges clearly from the comparison of the two systems as an intrinsic fingerprint of their correlations. Interestingly, these different widths are in very good agreement with predictions of cluster DMFT calculations in \214, where they also lead to a pseudogap behavior \cite{MoutenetPRB18}. We conclude that both \214 and \327 are suitable systems to study this physics in more details.

\section{Acknowledgements}
We thank M. Civelli and F. Bert for useful discussions. We thank ANR \lq\lq{}SOCRATE\rq\rq{} (ANR-15-CE30-0009-01), the Universit\'e Paris-Sud (\lq\lq{}PMP MRM Grant\rq\rq{}) and the Investissement d\rq{}Avenir LabEx PALM (ANR-10-LABX-0039-PALM) for financial support.

\bibliography{Iridates_biblio}

\vspace{0.5cm}

\section{Appendix}

\subsection{Sample characterization}
\begin{figure}[tbh]
\centering
\includegraphics[width=0.45\textwidth]{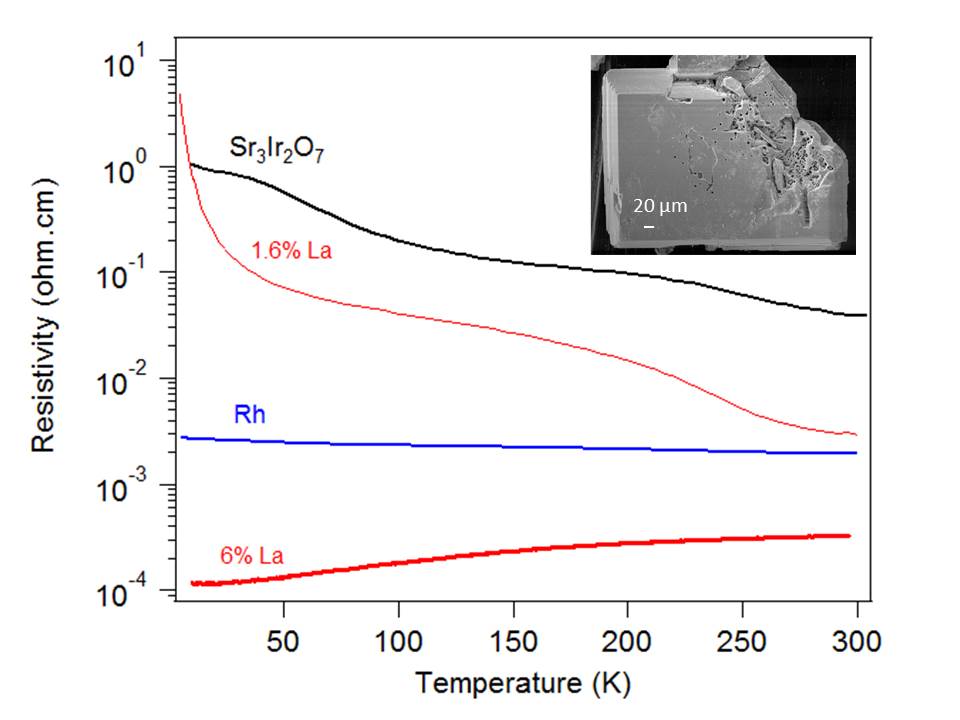}
\caption{Resistivity of the four samples used in this study. Inset : Image of \327 sample by electron microscope.}
\label{Fig_samples}
\end{figure}

In Fig. \ref{Fig_samples}, we show typical resistivity for samples used in this study. They were measured via standard four-wire measurements within a Quantum Design PPMS.

\subsection{Sketch of the band structure in S\lowercase{r}$_2$I\lowercase{r}O$_4$}
\begin{figure*}[tbh]
\centering
\includegraphics[width=0.85\textwidth]{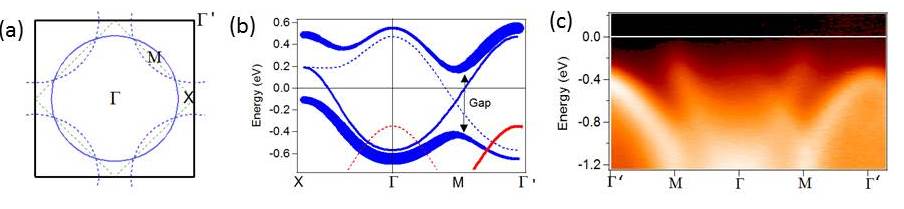}
\caption{(a) Sketch of the Fermi Surface expected for the J=1/2 band of \214. The black square is the 1Ir BZ and the dotted square the 2 Ir BZ. We call \lq\lq{}direct\rq\rq{} the bands of the 1Ir BZ (solid lines) and \lq\lq{}folded\rq\rq{} those obtained by folding in the 2Ir BZ (dotted lines). (b) Dispersion along X$\Gamma$M$\Gamma$ of the J=1/2 band (thin blue line) and its folded band (dotted blue line). Bold curves simulate the opening of a gap for an interaction at q=($\pi$,$\pi$), such as the AF order. The top of the J=3/2 band at $\Gamma$ is shown by red lines. (c) ARPES Energy-momentum plots of the band structure along $\Gamma$M in \214.} 
\label{Fig_Sup_214}
\end{figure*}

In Fig. \ref{Fig_Sup_214}, we show for reference a sketch of the band structure of \214, using only the J=1/2 band at the Fermi level (solid blue line). The expected Fermi Surface in DFT calculations is a circle. This band is folded with respect to the 2 Ir BZ boundaries, as shown as dotted line. The 2 inequivalent Ir arise due to rotations of the oxygen octahedra \cite{CrawfordPRB94}. The AF order gives rise to the same doubling of the unit cell. The direct and folded bands cross at the M point. 

In the case of AF order, a gap opens where the bands cross (arrow). The resulting dispersion is shown as thick blue lines. The size of the line is proportional to the expected ARPES spectral weight \cite{VoitScience00}. By contrast, in \327, this crossing is already gapped by interaction within the bilayer (see Fig. 1 and below), so that the effect of the AF order/correlations is only to enlarge this gap. 

Very similarly to \327, ARPES along $\Gamma$M (Fig. \ref{Fig_Sup_214}c) shows a band going up from $\Gamma$ to M, but the folded side form M to $\Gamma$' is missing.

\subsection{Band structure calculations for S\lowercase{r}$_3$I\lowercase{r}$_2$O$_7$}

Band structure calculations were done using the WIEN2k package \cite{Wien2k} and including spin-orbit coupling. Due to the complexity of the structure of \327 \cite{HoganPRB16} and to a systematic underestimation of the strength of SOC in LDA calculations \cite{LiuPRL08,ZhouPRX18}, calculations for \327 have often been done using tight-binding models \cite{CarterPRB13_327,DeLaTorrePhD}. Our results will be in qualitative agreements with these previous findings, but we use the actual experimental structure with space group \#68 (Ccca) to avoid using adjustable parameters. 

In Fig. \ref{Fig_Sup_327}(a), we highlight bands of J=1/2 and J=3/2 characters by blue and red colors, respectively. It is easy to see that the bands of \214 have split, by about 0.3eV at $\Gamma$. The J=3/2 band will be pushed lower in energy by a stronger SOC (below -0.2eV, see ref. \cite{MoreschiniPRB14,KingPRB13} and Fig. \ref{Fig_DatasPure}), so that we can focus on the J=1/2 band alone. We see that the J=1/2 bands interact with each other, opening a gap of structural origin in the black circles (100meV near X, 200meV near M). In this calculation, the small electron pockets at M are compensated by the J=3/2 hole pocket at $\Gamma$. If this band was pushed lower in energy, it would be compensated by narrow hole pockets at X, forming a semi-metallic structure presented in Fig. 1(a-b).  

In Fig. \ref{Fig_Sup_327}(b), we stabilize the magnetic state by adding an orbital potential U=2.7eV. This yields a magnetic moment M=0.28$\mu_B$, somewhat larger than in experiment \cite{DhitalPRB13}. The structure of J=1/2 into lower and upper band remains the same, but the gap between them (defined as in Fig. 1a) is enlarged to about 0.4eV.

\begin{figure*}[tbh]
\centering
\includegraphics[width=0.9\textwidth]{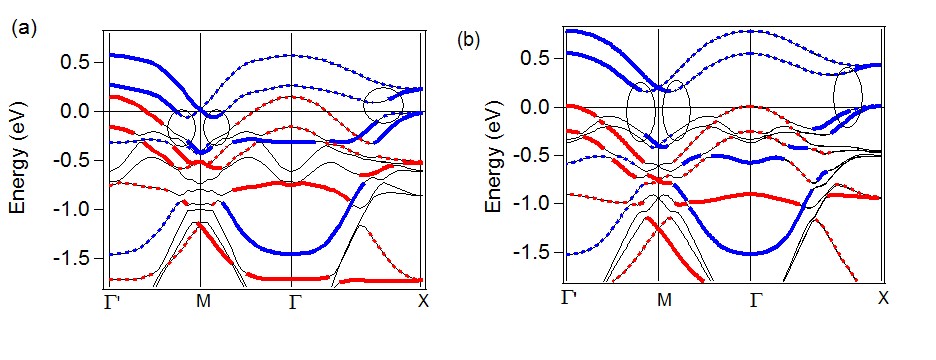}
\caption{(a) Black lines : band structure calculation for \327. The bands of dominant J=1/2 (J=3/2) character are emphasized by blue (red) lines. The folded bands are shown as dotted lines. (b) Calculation for the same compound, but with magnetism stabilized by U=2.7eV. } 
\label{Fig_Sup_327}
\end{figure*}
\end{document}